\documentclass[12pt]{article}

\usepackage{amssymb,amsthm}
\newcommand{\C}{{\mathbb C}}

\newcommand{\cH}{{\mathcal H}}

\newcommand\be{\begin{eqnarray}}
\newcommand\ee{\end{eqnarray}}

\begin{document}

\vspace{1.0cm}

\begin{center}
{\Large\textbf{Pleba\'nski Formulation of General Relativity:  }} 
\vskip 0.1cm
{\Large\textbf{A Practical Introduction}} 
\vspace{1.1cm}

\textbf{Kirill Krasnov}\footnote{kirill.krasnov@nottingham.ac.uk}  \\ \vspace{0.3cm}
\textit{School of Mathematical Sciences, University of Nottingham, Nottingham, NG7 2RD, UK}
\end{center}

\vspace{0.5cm}

\thispagestyle{empty}

\begin{center}
{\bf Abstract}
\end{center}
\vspace{-0.2cm}

We give a pedagogical introduction into an old, but 
unfortunately not commonly known formulation of GR in terms of self-dual two-forms 
due to in particular Jerzy Pleba\'nski. Our presentation is rather explicit in that we show how the familiar 
textbook solutions: Schwarzschild, Volkoff-Oppenheimer, as well as those describing the Newtonian limit, 
a gravitational wave and the homogeneous isotropic Universe can be obtained within this formalism. 
Our description shows how Pleba\'nski formulation gives quite an economical alternative to 
the usual metric and frame-based schemes for deriving Einstein equations. 

\newpage
\setcounter{page}{1}

\section{Plebanski formulation of general relativity}
\label{sec:Pleb}

The aim of this short paper is to give a description of Pleba\'nski self-dual formulation 
\cite{Plebanski:1977zz} of general relativity (GR) in a version that we found most suited for 
practical computations. Our presentation is very explicit, in
that the standard textbook solutions of GR are obtained. As we shall see, given an ansatz for the metric, 
Pleba\'nski formulation produces Einstein equations even more quickly than the already 
efficient tetrad method. In our opinion, the efficiency and beauty of this formulation
may warrant its inclusion in general relativity textbooks. 

Our convention for the signature is $(-,+,+,+)$. We start with a collection of historical remarks.

\subsection{Historical remarks}

The basic objects of Pleba\'nski's formulation of GR are self-dual two-forms. These objects
have appeared in the GR literature much before Pleba\'nski paper \cite{Plebanski:1977zz}. In fact,
the famous Petrov's classification \cite{Petrov:2000bs} of ``spaces defining gravitational fields'' 
already uses self-dual (and anti-self-dual) bivectors in a key way. Thus, the theorem proved by Petrov
states that the gravitational field (solution of vacuum Einstein equations) can be classified according to 
algebraic types of a complex symmetric $3\times 3$ matrix obtained as a complex
linear combination of the diagonal and off-diagonal blocks of the Riemann tensor viewed 
as a symmetric tensor in the linear space of bivectors. Self-dual (and anti-self-dual) bivectors
then naturally appear as principal bivectors of the Riemann tensor. A completely analogous but more modern treatment 
that forms the $3\times 3$ matrix in question as the complex linear combination of the 
``electric'' and ``magnetic'' parts of the Weyl tensor was given in \cite{Ehlers}. Again, the
self- and anti-self-dual bivectors are central in these considerations.  

It was then remarked in \cite{Taubes} that the $3\times 3$ complex matrix encoding the Weyl curvature 
can be computed directly, i.e. avoiding computing the Riemann curvature first. This can be done
by elementary operations of differentiation if one forms the self-dual complex linear
combinations from the components of the Levi-Civita tetrad-compatible connection. This encodes
24 real components of the Levi-Civita connection in 12 complex components of a self-dual connection.
The $3\times 3$ matrix of Weyl curvature components is then computed as the curvature of this
self-dual connection. 

The self-dual two-forms first appeared in a pioneering paper \cite{Cahen}.
This paper uses the null tetrad formalism and thus also provides a link to the spinor formalism of
Penrose \cite{Penrose:1960eq}. Thus, the spinor formalism combines the 24 real rotation coefficients into 12 complex
Newmann-Penrose spin coefficients \cite{Newman:1961qr}, which is similar to what happens in the self-dual formalism.  
The paper \cite{Cahen} for the first time
writes equations for the self-dual connection one-forms as those in terms of exterior derivatives of
the self-dual two-forms. It also clearly states that the isomorphism between the Lorentz group ${\rm SO}(1,3)$ and 
the complexified rotation group ${\rm SO}(3,\C)$ is what is at the root of the self-dual formalism.
Finally, the Einstein equations are very clearly stated in this paper as the condition that the curvature
of the self-dual connection is self-dual. Another exposition of the formalism for GR based
on differential forms and self-duality is that of \cite{Israel}.

Yet another presentation of the self-dual formalism for GR appeared in \cite{Brans:1974ry}. This reference is
very close in spirit to the exposition of the present paper. One important new point in this reference
is the emphasis it places on the role played by the Hodge duality operator, which is interpreted as
defining the complex structure in the space of two-forms. Similar to \cite{Cahen}, (vacuum) Einstein equations
are stated here as the condition that the curvature of the self-dual connection is self-dual. 

The self-dual formalism for general relativity was taken further by Pleba\'nski in \cite{Plebanski:1977zz}.
The paper cited uses spinor notations, but it can be easily translated into somewhat more easily readable ${\rm SO}(3)$
notations used in e.g. \cite{Brans:1974ry}. The main novelty of Pleba\'nski's work is that for the first time 
the main object of the theory is taken to be not a metric from whose tetrads the self-dual two-forms are constructed,
but rather a triple of two-forms satisfying certain additional equations. These equations guarantee
that the two-forms in question are obtained from tetrads, and thus provide a link to the usual metric formulation.
The paper \cite{Plebanski:1977zz} also gave a remarkably simple action principle realizing these ideas.
The basic dynamical field in this action is a triple of two-forms and no metric ever appears. Later
Ashtekar's new Hamiltonian formulation of general relativity \cite{Ashtekar:1987gu} was found \cite{Jacobson:1988yy}
to be just the phase space version of Pleba\'nski's theory.

Our presentation does not use the action principle \cite{Plebanski:1977zz}, 
and so all the technology employed here is already
available in earlier works \cite{Cahen} and \cite{Brans:1974ry}. Thus, it might perhaps be objectionable to name
the self-dual formulation of GR presented here after Pleba\'nski, since many earlier works contain the same
ideas. However, the self-dual formulation of general relativity has become firmly associated with the name of
Pleba\'nski in the quantum gravity community that uses the action principle of \cite{Plebanski:1977zz} as the
starting point for quantization of gravity, so this terminology is standard at least in some part of the GR community.  
Moreover, in our opinion it does take a considerable courage to propose a formulation of gravity in which the theory is
not about a metric. We believe that this justifies attributing the self-dual two-form
formulation to Jerzy Pleba\'nski. 

We proceed with a description of the key idea in the usual tensor notations.

\subsection{Einstein condition and the Hodge operator}

Given a spacetime metric $g_{\mu\nu}$ the condition that this metric is Einstein reads:  
$R_{\mu\nu}\sim g_{\mu\nu}$, where $R_{\mu\nu}:=R_{\mu\,\,\,\,\nu \rho}^{\,\,\,\rho}$
is the Ricci tensor, and the fact that the proportionality coefficient in this condition must be a constant
is implied by the (differential) Bianchi identity $\nabla^\mu G_{\mu\nu}=0$, where $\nabla_\mu$ is the
metric-compatible derivative operator and $G_{\mu\nu}=R_{\mu\nu}-(1/2)g_{\mu\nu} R$ is the
Einstein tensor. As usual, the quantity $R$ is the Ricci scalar $R:=R^\mu_\mu$, and all indices
are raised and lowered with the metric. The self-dual formulation of GR is based on the following
simple and well-known reformulation of the Einstein condition (vacuum Einstein equations) 
in terms of the Hodge operator. Thus, let us introduce the operation of Hodge dual that 
acts on bivectors (anti-symmetric rank two tensors) $A_{\mu\nu}$:
\be\label{Hodge}
A_{\mu\nu} \to {}^*A_{\mu\nu} = \frac{1}{2} \epsilon_{\mu\nu}^{\,\,\, \rho\sigma} A_{\rho\sigma},
\ee
where the quantity $\epsilon_{\mu\nu\rho\sigma}$ is the volume 4-form for the metric $g_{\mu\nu}$. 
The following elementary properties of the Hodge operator are easily verified: 
its square is minus one and it is invariant under conformal transformations of the
metric $g\to \Omega^2 g$.

Given the Riemann curvature tensor $R_{\mu\nu\rho\sigma}$
one can apply the Hodge operator to either the first or the second pair of indices:
\be
{}^* R_{\mu\nu\rho\sigma} := \frac{1}{2} \epsilon_{\mu\nu}^{\quad\mu'\nu'} R_{\mu'\nu'\rho\sigma}, 
\qquad R_{\mu\nu\rho\sigma}^* := \frac{1}{2} R_{\mu\nu\rho'\sigma'} \epsilon_{\quad\rho\sigma}^{\rho'\sigma'}.
\ee
It is a straightforward computation to check that the Einstein condition $R_{\mu\nu}\sim g_{\mu\nu}$
(together with the first Bianchi identity $R_{\mu[\nu\rho\sigma]}=0$) is equivalent to the condition
that the left and right Hodge duals of the Riemann tensor coincide:
\be\label{einst-cond}
R_{\mu\nu} \sim g_{\mu\nu} \Longleftrightarrow {}^* R_{\mu\nu\rho\sigma} = R_{\mu\nu\rho\sigma}^*.
\ee
Indeed, since $(*)^2=-1$ the last condition is equivalent to 
$R_{\mu\nu\rho\sigma} = - {}^*R_{\mu\nu\rho\sigma}^*$. Contracting a pair of indices
on the left-hand-side to produce the Ricci curvature, and expanding the product of two epsilons
on the right-hand-side in terms of products of the metric tensor, one finds 
$R_{\mu\nu} = (1/4) g_{\mu\nu} R$, which is the Einstein condition. 

To arrive at the self-dual formulation we need a slight additional reformulation of the condition
(\ref{einst-cond}). Thus, let us use the Hodge operator (\ref{Hodge}) to introduce projectors on the 
spaces of the so-called self- and anti-self-dual bivectors. These are bivectors that remain
essentially unchanged under the Hodge duality:
\be
{}^* A^{\mu\nu} = i A^{\mu\nu} \Longleftrightarrow ({\rm self-dual}), \qquad
{}^* A^{\mu\nu} = -i A^{\mu\nu} \Longleftrightarrow ({\rm anti-self-dual}).
\ee
The corresponding projectors are:
\be
P_{\mu\nu}^{\pm \,\, \rho\sigma} = \frac{1}{2} \left( I_{\mu\nu}^{\quad\rho\sigma} \pm \frac{1}{2i}
\epsilon_{\mu\nu}^{\quad\rho\sigma} \right),
\ee
where we have introduced the identity operator in the space of bivectors:
\be
I_{\mu\nu}^{\,\,\rho\sigma} := \delta_\mu^{[\rho} \delta_\nu^{\sigma]} = 
\frac{1}{2} \left( \delta_\mu^\rho \delta_\nu^\sigma 
- \delta_\mu^\sigma \delta_\nu^\rho \right).
\ee 

It is then easy to show that the condition (\ref{einst-cond}) can be rewritten using the
above projectors as follows:
\be\label{pleb-cond}
({\rm Ricci}) \sim ({\rm metric}) \Longleftrightarrow P^- R P^+ = 0,
\ee
where the anti-self-dual projector is applied on the left and the self-dual 
projector is applied on the right. Another convenient way to state this is to say that the
Einstein condition is equivalent to the statement that the self-dual part of the 
Riemann curvature with respect to the second pair of indices is self-dual with
respect to the first pair as well. As we have already described in the previous subsection,
this reformulation of the Einstein condition has been known for a long time, see e.g. 
\cite{Cahen} and \cite{Brans:1974ry}. We also note that in the mathematics literature, see
e.g. Chapter 13 of \cite{Besse}, this is
known as the Atiyah-Hitchin-Singer theorem \cite{Atiyah:1978wi}.
A proof of equivalence of (\ref{pleb-cond}) to (\ref{einst-cond}) is an 
elementary exercise using the definition of the projector operators. 
We can now use the observation (\ref{pleb-cond}) as the basis
of a new formulation of general relativity.

\subsection{Pleba\'nski self-dual formulation: pure gravity}

Let us present the self-dual formulation as a concrete recipe for obtaining Einstein
equations starting from a metric. We will then explain why this recipe is equivalent
to (\ref{pleb-cond}). Our description is quite analogous to that in \cite{Brans:1974ry}.
Thus, this reference already emphasizes the key fact that in the self-dual method the number
of equations to be solved is halved as compared to the tetrad method.

The first step of this formulation is to find a tetrad for the spacetime metric. Thus, one
represents the metric as: $ds^2= \theta^I \otimes \theta^J \eta_{IJ}$, where $\theta^I$ are
the tetrad one-forms, $I=0,1,2,3$ are ``internal'' indices, and $\eta_{IJ}$ is the Minkowski metric. 
The next step is to construct 
two-forms $\theta^I\wedge \theta^J$ and then take their self-dual parts with respect
to the indices $IJ$. More concretely, one introduces an arbitrary time plus space split
of the internal indices $I=(0,i), i=1,2,3$, and considers the following triple of two-forms:
\be\label{sigma}
\Sigma^i := i \theta^0 \wedge \theta^i - \frac{1}{2} 
\epsilon^{ijk} \theta^{j} \wedge \theta^{k}.
\ee
As is not hard to check, the two-forms $\Sigma^i$ satisfy:
\be\label{simpl}
\frac{i}{2} \Sigma^i \wedge \Sigma^j = \delta^{ij} \sqrt{-g}\, d^4x, \\
\label{reality} 
\Sigma^i \wedge \bar{\Sigma}^j = 0,
\ee
where the anti-self-dual forms $\bar{\Sigma}^i$ are given by:
\be\label{bar-sigma}
\bar{\Sigma}^i := i \theta^0 \wedge \theta^i + \frac{1}{2} 
\epsilon^{ijk} \theta^{j} \wedge \theta^{k}.
\ee

The next step, after the two-forms $\Sigma^i$ are written down, is to find
a connection $A^i$ that is ``compatible'' with the triple $\Sigma^i$, i.e.
the connection that satisfies:
\be\label{comp}
d\Sigma^i + \epsilon^{ijk} A^j \wedge \Sigma^k = 0.
\ee
This is an algebraic equation for the components of the connection $A^i$
that has a unique solution. It is not hard to write down an explicit
expression for $A^i$ in terms of the derivatives $d\Sigma^i$, 
but in practice it is easier to solve the equation (\ref{comp}) for each problem at hand 
by expanding it in the coordinate three-forms. We shall denote the $\Sigma$-compatible connection
by $A_\Sigma$. It is not hard to verify that $A_\Sigma$ is just the self-dual part
of the tetrad-compatible $d\theta^I + \omega^I_J \wedge \theta^J = 0$ 
connection $\omega^{IJ}$, i.e., $A^i=i\omega^{0i}-(1/2)\epsilon^{ijk}\omega^{jk}$. 
This fact will be important below for our explanation of the self-dual method from the 
point of view of (\ref{pleb-cond}), but it is not essential if one just want to follow the method. 

The final step is to compute the curvature 
\be
F^i = dA^i + \frac{1}{2} \epsilon^{ijk} A^j \wedge A^k
\ee
of the connection $A_\Sigma$. It is not hard to show that this is just the self-dual part
of the curvature $F^{IJ}(\omega)$ of the tetrad-compatible spin-connection $\omega^{IJ}$,
but, once again, this is not essential if one just wants to follow the method. 
Curvature is a two-form, and can be split in
the basis of self-dual $\Sigma^i$ and anti-self-dual $\bar{\Sigma}^i$ two-forms.
Thus, one can always write:
\be
F^i(A_\Sigma) = F^{ij} \Sigma^j + \bar{F}^{ij} \bar{\Sigma}^j.
\ee
The matrices $F^{ij}, \bar{F}^{ij}$ are the main quantities of interest
for the Einstein equations can be written quite easily as the following ten conditions:
\be\label{einst}
{\rm Tr}(F) = - \Lambda, \qquad \bar{F}^{ij} = 0,
\ee
where $\Lambda$ is the cosmological constant. As a bonus of this method, one not
only obtains Einstein equations, but also automatically
gets an expression for the part of the curvature not constrained by Einstein
equations  - the Weyl curvature. Thus, the self-dual part $\Psi^{ij}$ of the Weyl curvature
tensor is given by the tracefree part of the matrix $F^{ij}$: $\Psi^{ij}=(F^{ij})_{tf}$.

The described method of obtaining Einstein equations is quite efficient
for practical computations. The steps one has to take to arrive at (\ref{einst})
are similar to the steps one makes in the tetrad formulation. The advantage of
working with two-forms $\Sigma^i$ instead of tetrads $\theta^I$ is that one
has half equations to deal with at intermediate steps, at the expense of
all quantities becoming complex. We shall see the power of this method below
when we use it to obtain the standard solutions of GR.

An explanation of this method from first principles, i.e. from equation (\ref{pleb-cond})
is as follows. As we have already said, the $\Sigma$-compatible connection $A^i$
turns out to be just the self-dual part of the tetrad-compatible spin connection
$\omega^{IJ}$. This means that the curvature $F^i(A_\Sigma)$ is just the
self-dual part of the curvature tensor $R^{IJ}(\omega)$. Now the second 
equation in (\ref{einst}) just says that the curvature $F^i(A_\Sigma)$ is
self-dual as a two-form. However, because of the noted relation between
$F^i$ and $R^{IJ}$ this is equivalent to the condition that the self-dual
part of $R^{IJ}$ is self-dual as a two-form, which is just the condition
(\ref{pleb-cond}). The first equation in (\ref{einst}) can be shown
to be a consequence of the Bianchi identity $D_A F^i=0$, analogous
to how the proportionality coefficient in the Einstein condition
$R_{\mu\nu}\sim g_{\mu\nu}$ is shown to be a constant related to
the cosmological constant. For more discussion on abstract aspects of 
Pleba\'nski self-dual formulation of GR the reader may consult \cite{Cahen}, 
\cite{Brans:1974ry} and also \cite{Capovilla:1991qb}.

\subsection{Coupling to matter}

In the previous subsection we have described the vacuum theory. We now need
to extend it to the case when a non-zero stress-energy tensor
is present. This is not hard to do in full generality, but does not appear to have been spelled out
in the literature. Only the case of Einstein-Maxwell theory is commented upon in \cite{Cahen}.

To describe the coupling to matter it is necessary to split the stress-energy
tensor of matter $T_{\mu\nu}$ into its trace $T=T^\mu_\mu$ and the tracefree 
$\tilde{T}_{\mu\nu}=T_{\mu\nu}-(1/4)g_{\mu\nu}T$ parts, 
and then form out of the tracefree part a $3\times 3$ (complex) matrix
\be
T^{ij}= \tilde{T}_{\mu}^\rho \Sigma^i_{\nu\rho} \bar{\Sigma}^{j\,\mu\nu},
\ee
where, as before $\Sigma^i, \bar{\Sigma}^i$ are the self- and anti-self-dual forms (\ref{sigma}) and
(\ref{bar-sigma}). The non-vacuum Einstein equations (\ref{einst}) then take the 
following simple form:
\be\label{eqs*}
{\rm Tr}(F) = - \Lambda - 2\pi G T, \qquad \bar{F}^{ij}  = - 2\pi G T^{ij}.
\ee

For the ideal fluid, which is what matters for most practical applications, we have
$T_{\mu\nu}=(\rho+P) u_\mu u_\nu + P g_{\mu\nu}$, where $\rho, P$ are the energy and pressure densities
correspondingly, and $u_\mu$ is the 4-velocity vector. The trace of the stress-energy tensor is 
given by $T=(3P-\rho)$, the tracefree part $\tilde{T}_{\mu\nu}=(\rho+P)(u_\mu u_\nu + (1/4)g_{\mu\nu})$
and so we get:
\be
T^{ij} = (\rho+P) \left( \frac{\delta^{ij}}{1-|u|^2} - 2i\epsilon^{ijk} \frac{u^k}{\sqrt{1-|u|^2}}\right),
\ee
where $u^k$ is just the spatial component of the normalized $u_\mu u^\mu =-1$ velocity 4-vector: 
\be
u_\mu=\frac{1}{\sqrt{1-|u|^2}} ((dt)_\mu+u^i(dx^i)_\mu).
\ee 
In these two formulas $|u|^2=u^i u^i$, with, as usual, a sum over the repeated index $i$ implied.
Other types of matter can be described similarly.

We are now ready to study some simple solutions of GR using the formulation
described. We start with the graviton solution of the linearized theory. We shall
set the cosmological constant $\Lambda$ to zero from now on.

\section{Gravitational wave solution}

The gravitational wave is a solution of linearized around Minkowski spacetime vacuum
field equations. The background two-forms are given by:
\be\label{Sigma-Mink}
\Sigma^i_0 = i dt\wedge dx^i - \frac{1}{2}\epsilon^{ijk} dx^j\wedge dx^k.
\ee
In Plebanski formulation the graviton is described as a perturbation:
\be
\Sigma^i = \Sigma^i_0 + \delta \Sigma^i.
\ee
The perturbation two-forms $\delta\Sigma^i$ can then be decomposed into the background two-forms:
\be\label{perturb}
\delta \Sigma^i = b^{ij} \Sigma^j_0 + \bar{b}^{ij} \bar{\Sigma}^j_0.
\ee
In terms of the matrices $b^{ij}, \bar{b}^{ij}$ the reality conditions (\ref{reality}) become
\be\label{perturb-cond}
b^{ij} = b\delta^{ij} + \omega^{ij}, \qquad \bar{b}^{ij} = (\bar{b}^{ij})^*,
\ee
where $b$ is a real scalar and $\omega^{ij}$ is an arbitrary (complex) anti-symmetric
matrix. The second equation implies that the matrix of quantities $\bar{b}^{ij}$ is real. 

The representation (\ref{perturb}) of a general perturbation immediately provides
a convenient classification of perturbations. Indeed, the matrix $\bar{b}^{ij}$ can
be split into its symmetric traceless, anti-symmetric, and the trace parts, while
the matrix $\omega^{ij}$ is already anti-symmetric. Then the transverse part of an 
symmetric tracefree matrix describes a tensor mode, while the transverse part
of an anti-symmetric matrix describes a vector one. For example, perturbations of
the tensor type that correspond to gravitational waves are given by:
\be\label{perturb-grav}
\delta_{tens} \Sigma^i = h^{ij} \bar{\Sigma}^j,
\ee
where $h^{ij}$ is a symmetric, traceless, transverse $h^{ij}_{,i}=0$ three by three
matrix. 

Perturbations of the scalar and vector types can also be considered but the
modes they describe are non-propagating. Further, some of these modes
are gauge. For example, the modes described by $\omega^{ij}$ part of the
perturbation are just the gauge modes corresponding to availability of
(complexified) ${\rm SO}(3)$ transformations in this framework. Other modes
are gauge for the infinitesimal diffeomorphisms, which in this framework are described by:
\be\label{diffeo}
\delta_\xi \Sigma^i = {\mathcal L}_\xi \Sigma^i = d \iota_\xi \Sigma^i,
\ee
where ${\mathcal L}_\xi$ is the Lie derivative along an infinitesimal vector
field $\xi$ and $\iota_\xi$ is the operation of the interior product a vector
field with a form. It is quite easy to compute (\ref{diffeo}) and see which
modes can be set to zero by the diffeomorphisms. One finds
that, for example, the gauge-fixed perturbation of the scalar type is described by:
\be
\delta_{scal} \Sigma^i = \phi \Sigma^i + \psi \bar{\Sigma}^i,
\ee
where $\phi,\psi$ are related to the usual Newtonian potentials $\Phi,\Psi$ in an
elementary way. We shall further consider scalar perturbations below when
we study the Newtonian limit.

Having determined the form of the perturbation that we would like to study, it is
elementary to obtain the linearized Einstein equations. Thus, the first step
is to find the linearized connection $\delta A^i$ such that: 
$d\delta \Sigma^i + \epsilon^{ijk} \delta A^j \wedge \Sigma^k = 0$. This is
a simple exercise in algebra with the result being:
\be
\delta A^i = \left( - h^{ik}_{,l} \epsilon^{jkl} + i(h^{ij})'\right) dx^j.
\ee
Here the prime denotes the time derivative. The corresponding linearized curvature is also 
easy to compute. We decompose the answer into the self- and anti-self-dual background forms:
\be
d\delta A^i = \frac{1}{2} (h^{ij})'' (\Sigma^j + \bar{\Sigma}^j) + \frac{1}{2} \Delta h^{ij}
 (\Sigma^j - \bar{\Sigma}^j) + i (h^{ik}_{,l})'\epsilon^{jkl} \Sigma^j.
\ee
Now setting to zero the anti-self-dual terms on the right hand-side gives
the Einstein equations:
\be
 (h^{ij})'' - \Delta h^{ij} = \Box h^{ij}= 0,
\ee
whose solutions are plane waves. As a bonus, we also get an expression
for the matrix of the components of the self-dual part of the Weyl curvature of the 
linearized solution describing the gravitational wave:
\be
\Psi^{ij}=\frac{1}{2} \left( (h^{ij})'' + \Delta h^{ij} + 2i (h^{ik}_{,l})'\epsilon^{jkl}\right),
\ee
which, as is easy to check, is symmetric and traceless.

\section{Schwarzschild solution}

In this case one is still interested in vacuum field equations.
We start from the usual expression for a spherically-symmetric metric:
\be
ds^2 = -f^2(r) dt^2 +g^2(r) dr^2 + r^2 d\Omega^2,
\ee
where $d\Omega^2$ is the usual metric on the unit sphere. A tetrad one-form is given by:
\be\label{ss-tetrad}
e^t=f(r)dt, \qquad e^r=g(r)dr, \qquad 
e^\theta=r d\theta, \qquad e^\phi = r\sin(\theta) d\phi.
\ee
A convenient set of self-dual combinations (\ref{sigma}) is then:
\be\label{sch-B}
\Sigma^1=i e^t\wedge e^r - e^\theta\wedge e^\phi, \qquad
\Sigma^2=i e^t\wedge e^\theta - e^\phi\wedge e^r, \qquad
\Sigma^3=i e^t\wedge e^\phi - e^r\wedge e^\theta.
\ee

We now have to find the associated $\mathfrak{su}(2)$ connection $A^i$.
One decomposes the connection into coordinate one-forms, and then
solves the arising system of linear equations for the coefficients.
This gives:
\be
A^1= \frac{if'}{g}dt+ \cos(\theta)d\phi, \qquad A^2 = -\frac{\sin(\theta)d\phi}{g}, \qquad
A^3 = \frac{d\theta}{g},
\ee
where, as usual, prime denotes the derivative with respect to the $r$ coordinate.
It is not hard to verify that the above connection solves
(\ref{comp}).

In practice finding the connection is the most time consuming task. After this
is done, it only remains to compute the curvature. This is a simple exercise
in differentiation. One gets:
\be\nonumber
F^1 = dA^1 + A^2\wedge A^3 = -\left(\frac{if'}{g}\right)'dt\wedge dr 
- \left(1-\frac{1}{g^2}\right)\sin(\theta)d\theta\wedge d\phi, \\ \label{sch-F}
F^2 = dA^2 + A^3\wedge A^1 = - \frac{g'}{g^2}\sin(\theta) d\phi\wedge dr 
- \frac{if'}{g^2}dt\wedge d\theta, \\ \nonumber
F^3=dA^3+A^1\wedge A^2 = - \frac{g'}{g^2}dr\wedge d\theta 
- \frac{if'}{g^2}\sin(\theta)dt\wedge d\phi.
\ee

In the final step of extracting Einstein equations one simply has to
express the coordinate two-forms appearing in (\ref{sch-F})
in terms of the self- and anti-self-dual forms $\bar{\Sigma}^i=-(\Sigma^i)^*$. Thus, our final
expression for the curvature components is:
\be\nonumber
F^1 = -\frac{1}{2fg}\left(\frac{f'}{g}\right)'(\bar{\Sigma}^1+\Sigma^1)
- \frac{1}{2r^2}\left(1-\frac{1}{g^2}\right)(\bar{\Sigma}^1-\Sigma^1), \\
F^2 = - \frac{1}{2g^2r}\left( \frac{g'}{g}(\bar{\Sigma}^2-\Sigma^2)
+\frac{f'}{f}(\bar{\Sigma}^2+\Sigma^2) \right), 
\\ \nonumber
F^3=-\frac{1}{2g^2r}\left( \frac{g'}{g}(\bar{\Sigma}^3-\Sigma^3)
+\frac{f'}{f}(\bar{\Sigma}^3+\Sigma^3) \right).
\ee

After one has an expression for the curvature in terms of the basis
two-forms $\Sigma^i, \bar{\Sigma}^i$ one can immediately read off Einstein equations. First,
one should equate to zero all the anti-self-dual components of the curvature.
This immediately gives (from $F^2, F^3$):
\be\label{sch-1}
\frac{f'}{f}+\frac{g'}{g}=0,
\ee
whose solution is the familiar $f=1/g$. The equation coming from the anti-self-dual part of $F^3$ reads:
\be\label{sch-2}
\frac{1}{fg}\left( \frac{f'}{g}\right)' + \frac{1}{r^2}\left(1-\frac{1}{g^2}\right)=0.
\ee
The other equation, namely the condition that the trace part of the self-dual
part of the curvature is zero reads, with (\ref{sch-1}), (\ref{sch-2}) taken into account,
after some simple rewriting:
\be\label{sch-3}
\left( \frac{1}{g^2} \right)' = \frac{1}{r}\left(1-\frac{1}{g^2}\right),
\ee
which immediately gives the famous:
\be\label{sch-g}
\frac{1}{g^2} = 1 - \frac{r_+}{r},
\ee
which also solves (\ref{sch-2}), as is not hard to see.
The sign of the integration constant here is chosen so that the spacetime obtained
is that of positive total gravitational mass. The integration constant itself
is chosen so that $r=r_+$ is the place where $g^{-2}(r)$ vanishes.

We would like to emphasize how much more thought-economizing the above derivation
was as compared to the usual (e.g. tetrad-based) derivation given in all the
textbooks. Even the steps leading to the computation of the curvature involve
just the half of operations needed in the tetrad scheme (in the above 
spherically-symmetric case the economy is probably not as good, as a
large number of coefficients is anyway zero). However, what really is
most economical about Plebanski formulation is that one can immediately
read off the equations from the expression for the curvature, unlike
in the usual tetrad-based scheme. Indeed, in the tetrad scheme one first has to form
the Ricci tensor, and only after that write Einstein equations. 
In this last step one has to raise and lower indices, which introduces
some tricky minus signs, making it a non-trivial exercise to get the
right final expressions. In the case of Pleba\'nski formulation this
last step is absent altogether, the manipulations leading to 
Einstein equations are as algorithmic as all the previous ones,
making it much harder to make a mistake. 

\section{Volkoff-Oppenheimer solution}

In this section we switch on the stress-energy of matter. We have computed the curvature components in the previous 
section. The equations that one obtains from the anti-self-dual components of the curvature
are:
\be\label{volk-1}
\frac{1}{2 g^2 r}\left(\frac{g'}{g}+\frac{f'}{f}\right)=2\pi G (\rho+P),
\\ \label{volk-2}
\frac{1}{2fg}\left(\frac{f'}{g}\right)' + \frac{1}{2r^2}\left( 1-\frac{1}{g^2}\right)
= 2\pi G(\rho+P).
\ee
The equation one obtains from the trace of the self-dual part is given by:
\be\label{volk-3}
-\frac{1}{g^2 r}\left(\frac{f'}{f}-\frac{g'}{g}\right)
-\frac{1}{2fg}\left(\frac{f'}{g}\right)' + \frac{1}{2r^2}\left( 1-\frac{1}{g^2}\right)
=2\pi G(\rho - 3P).
\ee
Let us now take twice the equation (\ref{volk-1}) summed with (\ref{volk-2}), and subtract the
result from the equation (\ref{volk-3}). We get, after some simple rewriting:
\be\label{volkoff-eqn}
\frac{1}{r}\left( 1- \frac{1}{g^2}\right)'+\frac{1}{r^2}\left( 1- \frac{1}{g^2}\right) = 8\pi G \rho,
\ee
which is the non-zero $\rho$ generalization of the equation (\ref{sch-3}). It immediately
gives:
\be\label{volk-g}
g^{-2}(r) = 1- \frac{2m}{r}, \qquad m(r) = \int_0^r 4\pi G \rho r^2.
\ee
The function $f$ can then be found from e.g. equation (\ref{volk-1}). Substituting
(\ref{volk-g}) we get:
\be
\frac{f'}{f} = \frac{m+4\pi G P r^3}{r(r-2m)},
\ee
integrating which we get $f$. This ODE should be solved with the ``boundary'' condition
$f^2(R)=1-2M/R$, where $M, R$ are the total mass and radius of the spherical object in question.

\section{Newtonian limit}

Let us now consider the case of small (static) perturbations around the Minkowski
spacetime background (\ref{Sigma-Mink}) described by two scalar potentials $\phi,\psi$:
\be\label{newt-perturb}
\delta B^i = \phi \Sigma^i + \psi \bar{\Sigma}^i.
\ee
Here we assume the functions $\phi,\psi$ to be only those of the
spatial coordinates (time independent). It can be easily 
shown that all other scalar perturbations can be set to zero by a diffeomorphism
and an ${\rm SO}(3)$ rotation. It is easy to
check that (\ref{newt-perturb}) corresponds to the following perturbation
of the metric:
\be
ds^2 = -(1+2\Phi)dt^2 + (1-2\Psi)\sum_i dx_i^2,
\ee
with
\be\label{pot-metr}
\Phi = \frac{\phi+3\psi}{2}, \qquad \Psi = \frac{\psi-\phi}{2}.
\ee

We now have to solve the linearized ``compatibility'' equation 
and obtain the linearized connection. We can solve for the $\phi$ and $\psi$ parts
of the perturbation separately. We get:
\be
A_\phi^i = \frac{i}{2}\phi_{,i} dt - \frac{1}{2}\epsilon^{ijk} \phi_{,k} dx^j,
\qquad
A_\psi^i = \frac{3i}{2}\psi_{,i} dt + \frac{1}{2}\epsilon^{ijk} \psi_{,k} dx^j.
\ee
The corresponding linearized curvatures are:
\be
dA_\phi^i = -\frac{1}{4}\Delta\phi \Sigma^i + 
\frac{1}{4}(-2\phi_{,ij}+\delta^{ij} \Delta \phi) \bar{\Sigma}^j, \\ \nonumber
dA_\psi^i = -\frac{1}{4}(4\psi_{,ij}-\delta^{ij}\Delta\psi) \Sigma^j -
\frac{1}{4}(2\psi_{,ij}+\delta^{ij}\Delta \psi)\bar{\Sigma}^j.
\ee

Let us now analyze the Plebanski equations (\ref{eqs*}) with zero pressure $P=0$ and velocity $u^i=0$. 
First,  the anti-self-dual part on the right-hand-side is diagonal, which requires of the 
off-diagonal components to vanish:
\be
((\phi+\psi)_{,ij})_{tf}=0,
\ee
where $tf$ denotes the trace-free part. This implies $\phi+\psi=0$, or $\Psi=\Phi$. Both of
the diagonal components then give the Laplace equation, which in terms of the potentials
(\ref{pot-metr}) takes the familiar form:
\be
\Delta \Phi = 4\pi G \rho.
\ee
We also get for free an expression for the components of the Weyl tensor:
\be
\Psi^{ij} = \Phi^{,ij}- \frac{1}{3}\delta^{ij} \Delta\Phi,
\ee
which, we note, is real, as is typical of static backgrounds.

\section{Homogeneous isotropic Universe}

Using Plebanski formalism it is also quite easy to obtain the Friedman equations
describing the evolution of a homogeneous isotropic Universe. The corresponding
metric is given by:
\be\label{FRW}
ds^2 = a^2(\eta)\left(- d\eta^2 + \sum_i (dx^i)^2\right),
\ee
where $\eta$ is the conformal time.
The corresponding self- and anti-self-dual two-forms are given by:
\be\label{B-background}
\Sigma^i = a^2( i d\eta\wedge dx^i - \frac{1}{2} \epsilon^{ijk} dx^j \wedge dx^k ), \qquad
\bar{\Sigma}^i = a^2( i d\eta\wedge dx^i + \frac{1}{2} \epsilon^{ijk} dx^j \wedge dx^k ).
\ee
The connection $A^i$ compatible with the set of self-dual two-forms $\Sigma^i$, i.e.
satisfying (\ref{comp}), is given by:
\be\label{A-background}
A^i = i \cH dx^i,
\ee
where we have introduced:
\be
\cH := \frac{a'}{a}
\ee
and the prime denotes the derivative with respect to the conformal time. The
curvature $F^i = dA^i + (1/2)\epsilon^{ijk} A^j \wedge A^k$ of the connection
(\ref{A-background}) is given by:
\be\label{F-background}
F^i = \frac{1}{2a^2}  
(\cH'+\cH^2) \Sigma^i + \frac{1}{2a^2} (\cH'-\cH^2) \bar{\Sigma}^i.
\ee

From (\ref{F-background}), (\ref{eqs*}) we immediately get Einstein equations:
\be
\frac{1}{2a^2} (\cH'+\cH^2) = \frac{2\pi G}{3}(\rho-3P), \qquad
\frac{1}{2a^2} (\cH'-\cH^2) = -2\pi G (\rho +P).
\ee
These are more customarily known as their linear combinations: 
\be\label{Friedmann}
\cH^2 = \frac{8\pi G a^2 \rho}{3}, \qquad 2\cH' + \cH^2 = -8\pi G a^2 p.
\ee

\section{Discussion}

We would like to conclude this short paper by emphasizing once more how
computationally-efficient the Pleba\'nski self-dual formulation of general relativity
is as compared to other standard textbook methods such as that based on tetrads.
For this reason its inclusion in GR textbooks may be warranted, also
in view of its conceptual simplicity and beauty. 

What we have not discussed in this paper is if Pleba\'nski formulation can be not
just a tool for obtaining Einstein equations but also for solving them. To this end
we just note that the Bianchi identity takes in this formulation  a very simple
form:
\be\label{bianchi}
D_A \Psi^{ij}\wedge \Sigma^j =0,
\ee
where $\Psi^{ij}$ is the matrix of self-dual components of the Weyl curvature
tensor. In the case of the Schwarzschild solution the matrix $\Psi^{ij}$ is diagonal
$\Psi^{ij}=\beta(r){\rm diag}(2,-1,-1)$, where $\beta(r)$ is a function of the
radial coordinate. The Bianchi identity (\ref{bianchi}) then becomes a first
order differential equation for $\beta$, which immediately gives $\beta\sim 1/r^3$.
One then finds that one of the Einstein equations is an algebraic relation 
between $g(r)$ and $\beta(r)$, so the problem of solving the system of 
Einstein equations reduces in this case to that of solving (\ref{bianchi}).
This simple observation is relevant in situations other than
Schwarzschild, and in some cases serves as an efficient tool for obtaining solutions.

As we have already mentioned in the ``historical remarks'' subsection,  
the described here method is closely related to that
of Newman-Penrose \cite{Penrose:1985jw}, with the later being, loosely
speaking, the two-component spinor version of the self-dual two-form one. In addition to working
with self-dual quantities, and expressing all quantities in terms of spinors,
the Newman-Penrose formalism introduces and works with 
a doubly-null tetrad, which is quite powerful, but at the same 
time makes the Lorentz-covariance of the theory not manifest. In contrast, Lorentz
rotations of a tetrad in Pleba\'nski formalism are described very simply as 
(complexified) ${\rm SO}(3)$ rotations. It is also considerably more easy to work with spacetime
forms than with spinors. For these reasons Pleba\'nski method may be 
an convenient middle ground between the relatively cumbersome tetrad method
and a very powerful, but somewhat too relying on working in components 
Newman-Penrose formalism.

\section*{Acknowledgements} The author was supported by an EPSRC Advanced Fellowship. The author
is grateful to an anonymous referee for suggesting a list of references that grew into the
``historical remarks'' subsection.

\end{document}